\documentclass[conference]{IEEEtran}
\IEEEoverridecommandlockouts
\usepackage[top=0.75in, left=0.7in, right=0.7in, bottom=1.05in]{geometry}
\usepackage{amsmath,amssymb,amsfonts}
\usepackage{algorithmic}
\usepackage{graphicx}
\usepackage{textcomp}

\usepackage[table,xcdraw]{xcolor}

\usepackage{multirow}
\usepackage{ragged2e}
\def\BibTeX{{\rm B\kern-.05em{\sc i\kern-.025em b}\kern-.08em
    T\kern-.1667em\lower.7ex\hbox{E}\kern-.125emX}}

\begin{document}
\title{Network Digital Twin for Route Optimization in 5G/B5G Transport Slicing with \textit{What-If} Analysis}

\author{
	\IEEEauthorblockN{Rebecca Aben-Athar$^{\ast}$, Heitor Anglada$^{\ast}$, Lucas Costa$^{\ast}$, João Albuquerque$^{\ast}$, Abrahão Ferreira$^{\ast}$,\\ Cristiano Bonato Both$^{\dagger}$, Kleber Cardoso$^{\dag}$, Silvia Lins$^{\ddagger}$, Andrey Silva$^{\ddagger}$,\\Glauco Gonçalves$^{\ast}$, Ilan Correa$^{\ast}$, and Aldebaro Klautau$^{\ast}$}
    \IEEEauthorblockA{$^{\ast}$Federal University of Par\'{a}, Brazil. \\
	                   $^{\dagger}$University of Vale do Rio dos Sinos, Brazil. \\
                          $^{\dag}$Universidade Federal de Goi\'{a}s, Brazil,
                          $^{\ddagger}$Ericsson Research, Brazil,\\
					   E-mails: rebecca.athar@itec.ufpa.br,  cbboth@unisinos.br, kleber@inf.ufg.br, 
                        silvia.lins@ericsson.com\\
       \thanks{© 2025 IEEE.  Personal use of this material is permitted.  Permission from IEEE must be obtained for all other uses, in any current or future media, including reprinting/republishing this material for advertising or promotional purposes, creating new collective works, for resale or redistribution to servers or lists, or reuse of any copyrighted component of this work in other works.}
	}
}

\maketitle

\begin{abstract}
The advent of fifth-generation (5G) and Beyond 5G (B5G) networks introduces diverse service requirements, from ultra-low latency to high bandwidth, demanding dynamic monitoring and advanced solutions to ensure Quality of Service (QoS). The transport network - responsible for interconnecting the radio access network and core networks - will increasingly face challenges in efficiently managing complex traffic patterns. The Network Digital Twin (NDT) concept emerges as a promising solution for testing configurations and algorithms in a virtual network before real-world deployment. In this context, this work designs an experimental platform with NDT in a transport network domain, synchronizing with the virtual counterpart and a recommendation system for \textit{what-if} analysis, enabling intelligent decision-making for dynamic route optimization problems in 5G/B5G scenarios. Our NDT, composed of a Graph Neural Network (GNN), was evaluated across three different network topologies consisting of 8, 16, and 30 nodes. It achieved lower MAPE values for URLLC and eMBB slices, comparing latency predictions with actual latency after the solution implementation. These values indicate high accuracy, demonstrating the solution's effectiveness in generating precise insights into network performance if a particular solution were implemented.
\end{abstract}
    

\bibliographystyle{IEEEtran}
\section{Introduction }
\label{sec:introduction}

The rapid evolution of communication technologies, driven by the arrival of fifth-generation (5G) and  Beyond 5G (B5G) networks, has increased the demands for connectivity and real-time applications~\cite{ericsson_2024}. Consequently, telecommunication networks have become increasingly complex and heterogeneous due to the exponential growth of connected devices and the diverse services offered \cite{kadir_et_al.}. In this context, implementing network slicing \cite{ns} is crucial, allowing shared physical resources to be configured to meet different performance requirements, from ultra-low latency services such as autonomous vehicles and remote surgery \cite{10472343} to services that require high data rates, such as video streaming and Augmented Reality (AR) \cite{10233627}.

To support these diverse and demanding applications, a robust transport network is crucial, linking the Radio Access Network (RAN) to the Core and ensuring efficient data transmission between infrastructure components \cite{transport_net_5G}, thereby maintaining Quality of Service (QoS). However, traditional transport networks struggle to meet the varying demands of modern traffic, as routing typically follows predefined policies, often based on cost \cite{ospf}, without considering factors such as traffic heterogeneity, congestion, or real-time network conditions. Moreover, making network modifications, such as adding or removing links, implementing new routing policies, or creating a new slice, becomes increasingly challenging as ineffective decisions can result in congestion, packet loss, and high latency, ultimately degrading user experience and increasing operational costs.

The concept of  Digital Twin (DT) emerges as a promising solution for managing dynamic and complex networks. In the context of telecommunications, the  Network Digital Twin (NDT) enables real-time monitoring and automation by interactively mapping the target network in a virtual representation \cite{itu_digital_twin,etr_dt,network_DT}. This approach enables the emulation of scenarios, anticipating impacts of management modifications and optimizing resources.  In this context, several studies explore the application this concept in the transport network to optimize metrics such as latency, slice admission control, and routing policies \cite{ferriol, gnn_admission, raj}. However, there is still a gap in the literature on closed-loops that use the virtual counterpart to test new implementations and evaluate the reliability of the model in predicting network performance.

In this way, our work implements an flexible platform with NDT in the transport domain, enabling real-time network monitoring, using a recommendation system that tests different solutions in the virtual counterpart, allowing \textit{what-if} analysis for intelligent decisions before applying in the physical network, that can be applicable in different scenarios. For network modeling, we used Graph Neural Network (GNN), by overcoming traditional Deep Learning (DL) models due to their ability to generalize to topologies not seen in training and enable a deeper understanding of the intricate relationships between network topology, routing, and traffic patterns \cite{routenet-fermi}. In this work, we focus on route optimization for traffic with varying latency requirements, evaluating QoS degradation due to switch failures, mitigating Service Level Agreement (SLA) violations. This article aims to show the importance of NDT in providing insights into network performance if a particular solution were implemented, showing the impact of this technology on network management.

In summary, the contributions of this paper are as follows:
\begin{itemize}
    \item Development of a customizable NDT platform for 5G/B5G transport networks, integrating a recommendation system for \textit{what-if} analyses and route optimization using a surrogate Artificial Intelligence (AI) model.
    \item Compare the network performance predicted by GNN with the performance observed when the solution is implemented in the physical network. 
    \item Generate insights into how NDT performance varies with network scalability. 
\end{itemize}
The remaining sections of this paper are organized as follows: In Section \ref{sec:related_work}, we discuss previous work that addresses the usage of NDT technology in different network domains and face our solution with the literature. In Section III, we present our NDT solution and the \textit{what-if} analysis we are working on. Section \ref{sec:Prototype} describes our implementation. Section \ref{sec:simulations} shows the results of our experiments, followed by the paper's conclusion and future works in Section \ref{sec:conclusions}.

\section{Related work}
\label{sec:related_work}

\begin{figure*}[!ht]
    \centering
    \includegraphics[width=0.56\textwidth]{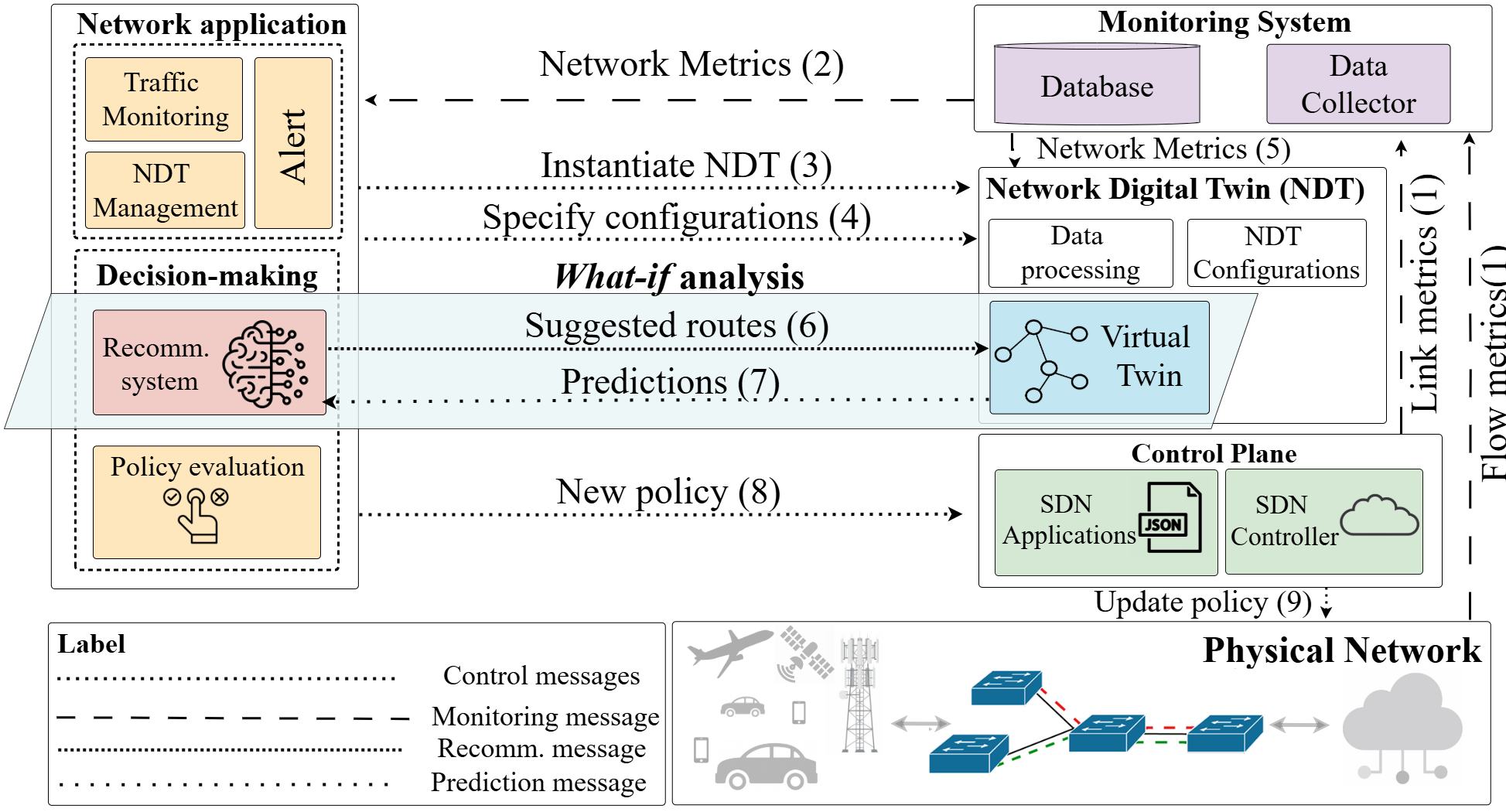}
    \caption{The figure above shows our architecture with NDT for transport networks. The figure's colors are distributed as follows: green for the Control Plane, blue for the virtual twin, purple for the Monitoring System, pink for the recommendation system, and yellow for the Control Functions of the Network Application layer.}
    \label{fig:architecture}
\end{figure*}

Network optimization and more informed decision-making have become increasingly critical with the expansion of connectivity and services enabled by 5G and B5G. In the transport network context, NDT is an important tool for network automation, enhancing traffic management, QoS metric analysis, and more efficient resource allocation. Consequently, various studies have been dedicated to developing a NDT in the domain.

In this sense, some works have focused on developing the virtual twin. Ferriol et al. \cite{ferriol} proposed the TwinNet, a GNN-based network model that can accurately estimate relevant SLA metrics for network optimization. C. Modesto et al. \cite{rebecca} used GNN to predicted QoS metrics (e.g., mean per-flow delay) in a real network dataset. Wang et al. \cite{slice_gnn} monitored end-to-end metrics of slices and used GNN to predict off-line traffic latencies in an 5G network slicing context. However, despite their effectiveness in specific tasks, these studies predominantly focused on modeling the virtual twin without considering its interaction with other NDT modules in a closed-loop and actions executed in the physical counterpart. Furthermore, the lack of an end-to-end test environment with dynamic traffic makes it difficult to understand the system behavior under adverse conditions and perform more accurate tests on the network. 

To meet this need, some studies approached the physical network. Messaoudi et al. \cite{gnn_admission} implemented an NDT, in the 5G context, using the surrogate model for an intelligent admission control problem, which decides whether to accept or reject new traffic, to ensure network requirements such as latency and packet loss. Chang et al. \cite{9943779} proposed a knowledge graph-enable intent-driven with NDT, focusing on the intent translation, enabling a better understanding of the user context and improving network policy generation. D. R. R. RAJ et al. \cite{raj} focused on evaluating the construction and the query efficiency of a NDT architecture for Software-Defined Networking (SDN), leverage by a Knowledge Graphs (KG) for data storage and modeling. M. Polverini et al. \cite{polverini} is close to our work by proposing a NDT comprising \textit{what-if} analysis, but for routing optimization inter-Autonomous Systems via Border Gateway Protocol (BGP). These previous work has shown promising results, with the NDT (or modules within an NDT) performing well on specific tasks. 

However, although these approaches explore ways to improve policy generation, routing, and user experience, they lack decision-making flexibility and are often limited to a single solution. This limitation can be a disadvantage in dynamic and complex environments, hindering valuable insights for new network implementation, which is a critical need to address the diversity of modern scenarios. 

Our work differs from others by proposing an experimental and customizable platform that integrates a recommendation system with the virtual counterpart. This system evaluates multiple solutions for different traffic types, performs \textit{what-if} analyses, and monitors network performance in real-time as these solutions are deployed in the physical network. This combination improves real-time adaptability, promotes insights, and enhances decision-making, addressing the limitations observed in more static models.

\section{NDT for transport network }

Our work aims to answer questions that facilitate better network decision-making. Among these questions are: \textit{"How accurate is the network performance prediction made by Virtual Twin compared to the actual performance after implementing the action on the Physical network?"} and \textit{"How does NDT performance vary as network size increases?"}. To address these questions, traffic with different latency requirements is used to evaluate the proposed architecture illustrated in Fig. \ref{fig:architecture}.

Initially, the Physical Network is continuously monitored through traffic metrics collected by the Data Collector and link-level metrics extracted by the SDN Controller (1). These metrics are stored in the Database and sent to the Network Application layer (2), where the network operator can monitor them. Under normal conditions, the transport network operates smoothly, with different traffic flows routed through strategic paths that meet the QoS requirements of each service, thus ensuring SLA compliance.

When a switch in use fails, the affected traffic suffers packet loss, which increases latency and compromises QoS. The Application layer detects that the requirements of the affected slices are not being met, triggering a control message that initiates the NDT (3), followed by the necessary configurations (4), such as specifying the type of virtual network representation to be used (e.g., emulation, surrogate AI model, etc.), and defining the use case to be addressed. At this stage, network metrics and topology data are retrieved from the Database and sent to NDT (5), where they are processed and provided as input for the virtual twin. At the same time, the recommendation system is activated to suggest new routes that best serve the affected traffic, using the same metrics and topology as a basis. These routes are sent to the virtual twin (6), which tests the suggestions, predicts the QoS metrics for each route, and sends the predictions to the Application layer (7). The virtual counterpart and the different routing algorithms in the recommendation system form the \textit{what-if} analysis, providing insights into how the network would behave if the suggested routes were implemented.

\begin{figure}[ht!]
    \centering
    \includegraphics[scale=0.4]{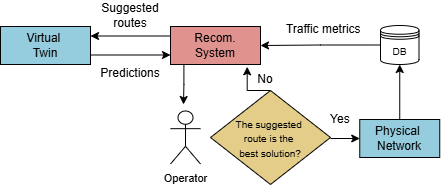}
    \caption{The representation of decision-making with the network operator options (accept or reject the solution) highlight in yellow.}
    \label{fig:what_if}
\end{figure}

The network operator selects the most appropriate solution to apply in the Physical Network based on the predictions generated by the virtual twin aiming to best meet the SLA of each service. Once the new policy is approved, the new route is sent to the SDN Applications (8), which act as translators, receiving the new route and transforming it into a network policy that is sent to the SDN Controller, which then updates the routing tables and implements the changes in the Physical counterpart (9). If the operator decides to reject the suggestions, the recommendation system generates new routes by repeating the process, as shown in Fig. \ref{fig:what_if}.

\label{sec:DT}

\section{Prototype }
\label{sec:Prototype}

In this session, we provide a detailed exploration of how each layer of our NDT was implemented in our prototype, explaining their workflows and the main strategies adopted to cope with the route optimization problem.

\subsection{Physical network}

The physical network topology is represented as a directed graph, where each node corresponds to a network switch. The architecture follows a pairwise manner and is defined by a 4-tuple graph $\mathcal{J} = (\mathcal{V}, \mathcal{E}, \textbf{r}, \textbf{s})$, where $\mathcal{V}$ is the set of nodes, $\mathcal{E}$ is the set of edges such that $\mathcal{E} \subseteq \mathcal{V} \times \mathcal{V}$, and $\textbf{r}$ and $\textbf{s}$ denote the features of the nodes and edges, respectively. Each link $(u, v)$ connects node $u$ to node $v$ with heterogeneous capacity. All network traffic originates from a single node and converges to another distant node, resulting in multiple possible route combinations between them.

In this work, we consider two service types: the enhanced Mobile Broadband (eMBB) slice, which has moderate latency constraints and bandwidth, and the Ultra-reliable Low-Latency Communications (URLLC) slice, which demands ultra-low latency and lower bandwidth. The specific requirements for each slice and the technical configurations of the physical network are detailed in Table~\ref{tab:phy}.

\begin{table}[htp!]
\centering
\caption{Technical configuration details} 
\scalebox{1.15}{\begin{tabular}{c|cc}
\hline
\rowcolor[HTML]{EFEFEF} 
Parameter & \multicolumn{2}{c}{\cellcolor[HTML]{EFEFEF}Value}  \\ \hline
\hline
Operating System  & \multicolumn{2}{c}{Ubuntu 20.04.4 LTS}  \\
\rowcolor[HTML]{EFEFEF} 
Software & \multicolumn{2}{c}{\cellcolor[HTML]{EFEFEF}\begin{tabular}[c]{@{}c@{}}ONOS 2.7-latest, Mininet\cite{mininet} 2.3.0, \\ Python 3.8.19\end{tabular}} \\
Protocols & \multicolumn{2}{c}{OpenVSwitch 2.13.8} \\
\rowcolor[HTML]{EFEFEF} 
Packet size   & \multicolumn{2}{c}{\cellcolor[HTML]{EFEFEF}125 bytes}  \\
\multirow{2}{*}{Bandwidth} & \multicolumn{1}{c|}{eMBB} & URLLC \\ \cline{2-3}
                           & \multicolumn{1}{>{\centering\arraybackslash}p{1.85cm}|}{10 Mbps} & 1 Mbps\\
\rowcolor[HTML]{EFEFEF} 
Max Latency & \multicolumn{1}{c|}{ \cellcolor[HTML]{EFEFEF} 10 ms} & 1 ms \\ \hline
\end{tabular}}
\label{tab:phy} 
\end{table}

We limited the traffic of both slices due to computational constraints, while maintaining the appropriate proportions. The solution and the results presented can be scaled and replicated for bandwidths of grater magnitude. Since all links have sufficient capacity to handle traffic from both slices simultaneously, congestion is not a concern. Thus, the focus is solely on the total delay associated with each path.

\subsection{Control plane}

The Control Plane is responsible for sending link metrics to the Database and implementing decisions in the physical network. When the operator approves a new route, it is sent from the Application layer to the Control Plane, where the SDN Applications framework processes it. 

The eMBB and URLLC slices are fed by two UDP streams. The Type of Service (ToS) field in the IP header is used to differentiate these slices, identifying the packet's QoS class. The SDN Application are implemented on top of the Open
Network Operating System (ONOS)\cite{onos}, the SDN Controller, exploring its Northbound REST API interface. In summary, the SDN Applications receive the path for the slice with its ToS, retrieve the switches and their interconnection ports, and map the path to the respective switch ports. The SDN Applications send ONOS's intents to the Controller, which updates the flow table at each switch through Openflow messages, rerouting traffic within the physical network.

\subsection{Monitoring system} \label{data_collector}
The Monitoring System plays a vital role in collecting, storing, and transmitting traffic metrics to the Application layers and the NDT when instantiated. The Data Collector gathers flow metrics for each slice directly from the physical network, and stores them in the Database. The Database also stores link metrics, such as capacity and load, which represent the maximum transmission rate a network link can support, obtained from the SDN controller.

\subsection{Network virtual twin}
\subsubsection*{Virtual twin}

For the virtual twin representation, we employed a GNN as a surrogate AI-based model - a neural network architecture (NN) specifically designed to learn from graph-structured data. This type of NN excels at tackling complex tasks by efficiently processing large-scale data and generalizing across unseen topologies, overcoming the limitations of conventional approaches \cite{routenet-fermi} using the so-called message-passing algorithm, which aggregates node features. In this context, nodes represent key network entities (e.g. switches), while edges correspond to the connections that facilitate traffic flow, collectively defining the network topology and optimizing data transmission.

A key aspect of message-passing in GNNs is how node features are updated during the learning process. Bronstein et al. \cite{brody2022} established a close relationship between attentional GNN models, such as Graph Attention Networks (GAT), with Message-Passing Neural Networks (MPNN). Specifically, they describe MPNN as a generalization of GAT models, where the primary distinction lies in how the \emph{update} function is defined in each model. Given this generalization, it is natural to leverage attention mechanisms by incorporating them into the \emph{update} function of the MPNN model, enhancing its ability to capture complex dependencies within the network topology.

Based on this, we used the GNN architecture proposed in \cite{rebecca}, which builds upon the Route-Net Fermi \cite{routenet-fermi} model. This architecture introduces a modification in the message-passing phase of the MPNN, enabling iterations over multiple stages and allowing the model to refine its representations progressively. In addition, it incorporates an attention mechanism layer to dynamically adjust the importance of the features. This was done by applying
\begin{equation}
\alpha_{uv} = \dfrac{\exp(\textbf{a}^T\text{LeakyReLU}( \mathbf{W}[\textbf{h}_u||\textbf{h}_v]))}{\sum_{k \in \mathcal{N}_u} \exp(\textbf{a}^T\text{LeakyReLU}( \mathbf{W}[\textbf{h}_u||\textbf{h}_k]))},
\label{eq:brody_attention_layer}   
\end{equation}

 \noindent where $\alpha_{uv}$ represents the normalized attention coefficient between node $u$ and $v$; \textbf{a} is the learnable weight vector and the $\textbf{W}$ is a learnable weight matrix; the $.^{T}$ is the transpose and the ($||$) is the concatenation operation, respectively, used in the aggregation of interest node features $\textbf{h}_u$ and its neighborhood node features $\textbf{h}_k$, such that $k \in \mathcal{N}_u$. Where $\mathcal{N}_u$ is the neighborhood of interest node $u$. 

The virtual twin was pre-trained with on a dataset\footnote{https://bnn.upc.edu/challenge/gnnet2023/dataset/} using a topology ranging from 5 to 8 nodes and tested on various topologies with different node counts,  demonstrating its generalization capability. The virtual twin inputs include:
\begin{enumerate}
    \item The flow metrics: $\textbf{r} \in \mathbb{R}^{n}$ where $n$ is the number of flow features, represented as the set $\mathcal{M} = \{R_1, R_2, \dots, R_n\}$, with each $R_i$ denoting a distinct feature. 
    \item The link metrics: defined as $\textbf{s} \in \mathbb{R}^{2}$ with the number $2$ representing the number of link features.
    \item The suggested route from the recommendation system: A subset of $\mathcal{J}$ represented by $\mathcal{P}$, where $\mathcal{P} = (\mathcal{V}_p, \mathcal{E}_p, \textbf{r}_p, \textbf{s}_p)$, where $\mathcal{V}_p \subseteq \mathcal{V}$ and $\mathcal{E}_p \subseteq \mathcal{E}$ correspond, respectively, to the subset of nodes and links that form the selected route.
\end{enumerate}

Thus, the input characteristics for each experiment are provided in Table \ref{tab:all_features}.

\begin{table}[htp]
\centering
\caption{Features used as input for the virtual twin} 
\scalebox{1.3}{\begin{tabular}{c|c|c}
\hline
\rowcolor[HTML]{EFEFEF} 
Features & Feature Role & Unit \\ \hline
\hline 
Average bandwidth & Flow & Mbits/$s$ \\
\rowcolor[HTML]{EFEFEF} 
Average packet size & Flow & Bytes\\
Average packet rate & Flow & pps \\
\rowcolor[HTML]{EFEFEF} 
Propagation delay & Flow & m$s$\\
Link capacity  &  Link & Mbits/$s$\\
\rowcolor[HTML]{EFEFEF} 
Link load & Link & ratio\\

\hline
\end{tabular}}
\label{tab:all_features} 
\end{table}

\subsection{Network applications}
\subsubsection*{Recommendation system}
An alert is sent to the recommendation system when an SLA violation is detected in the network. From now on, the recommendation system becomes active, suggesting alternative routes for the traffic that previously passed through the failed switch. To validated our proposed NDT, we chose two different methods for route recommendation.

\begin{itemize}
    \item Random: This method selects a route randomly among all available routes without considering traffic requirements.  
    \item AI-based model: We used Reinforcement Learning (RL) with the Proximal Policy Optimization (PPO) algorithm to identify the paths that best meet the requirements of each service in terms of latency.
\end{itemize}

\section{Performance evaluation}
\label{sec:simulations}

Our architecture was evaluated in a route optimization scenario, comparing network performance between NDT-based decisions and traditional methods, and analyzing the impact of this technology on network management in contexts where SLA is violated. We used three synthetically generated network datasets: the first, an 8-node topology with 15 links; the second, a 16-node topology with 27 links; and the third, a 30-node topology with 75 links. All links are unidirectional.

\begin{figure}[!ht]
   \centering
    \includegraphics[width=0.48\textwidth]{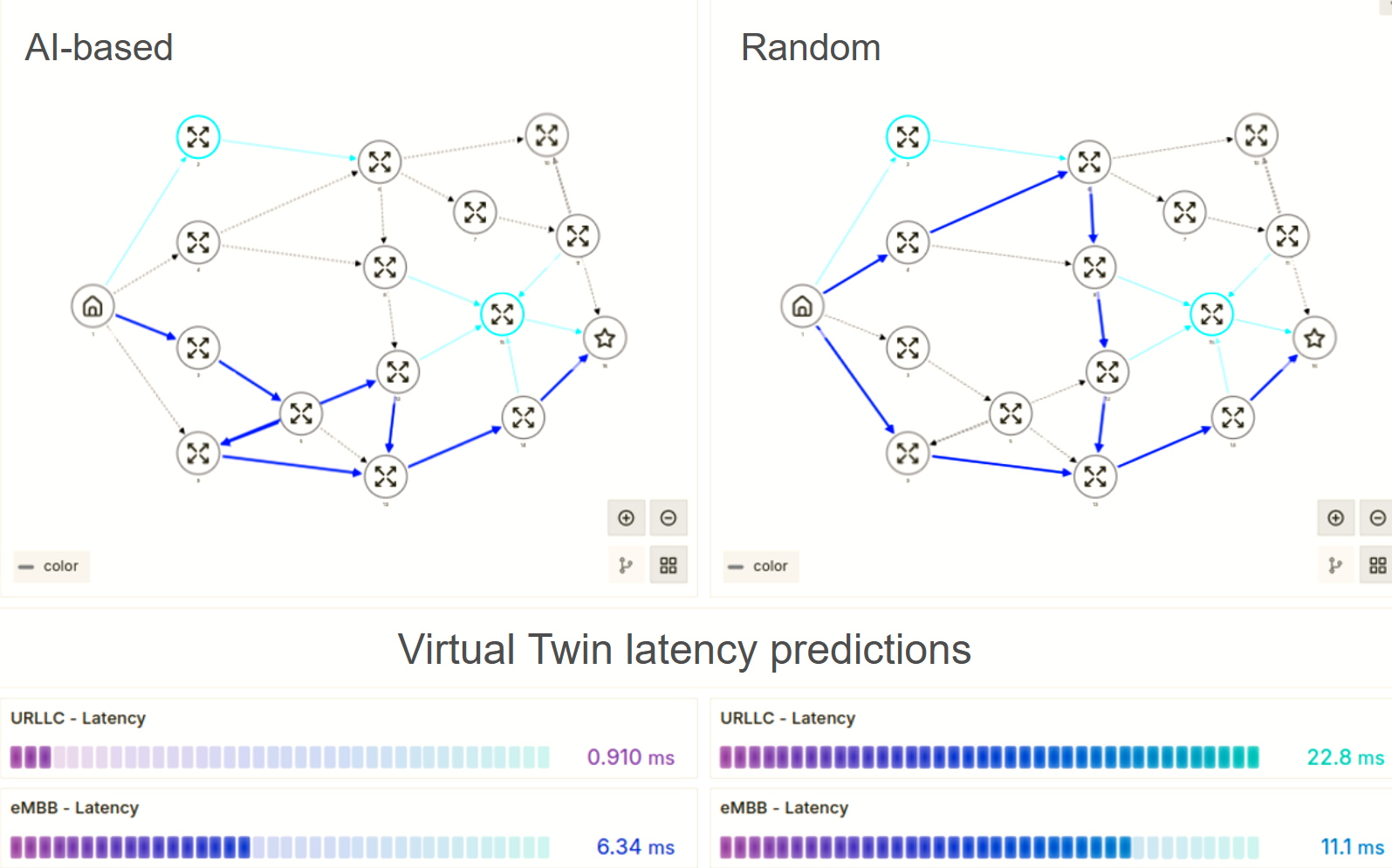}
    \caption{The suggested routes by the recommendation system, for the AI-based and Random solution, and the GNN latency predictions for each solution, composing the \textit{what-if} analysis, for the 16-node topology. The house and star symbol represent the source and destination switch, respectively.}
    \label{fig:dashboard}
\end{figure}

Fig. \ref{fig:dashboard} shows how NDT works for a 16-node topology. The image shows the routes suggested by the recommendation system – both for the AI-based and Random algorithm – for eMBB and URLLC slices and the virtual twin delay predictions for each solution, composing the \textit{what-if} analysis. The dark blue lines represent the path suggested by the routing algorithm, while the light blue lines indicate the path followed by the flow before the switch's failure occurred.

\begin{figure}[ht!]
    \centering
    \includegraphics[width=0.48\textwidth]{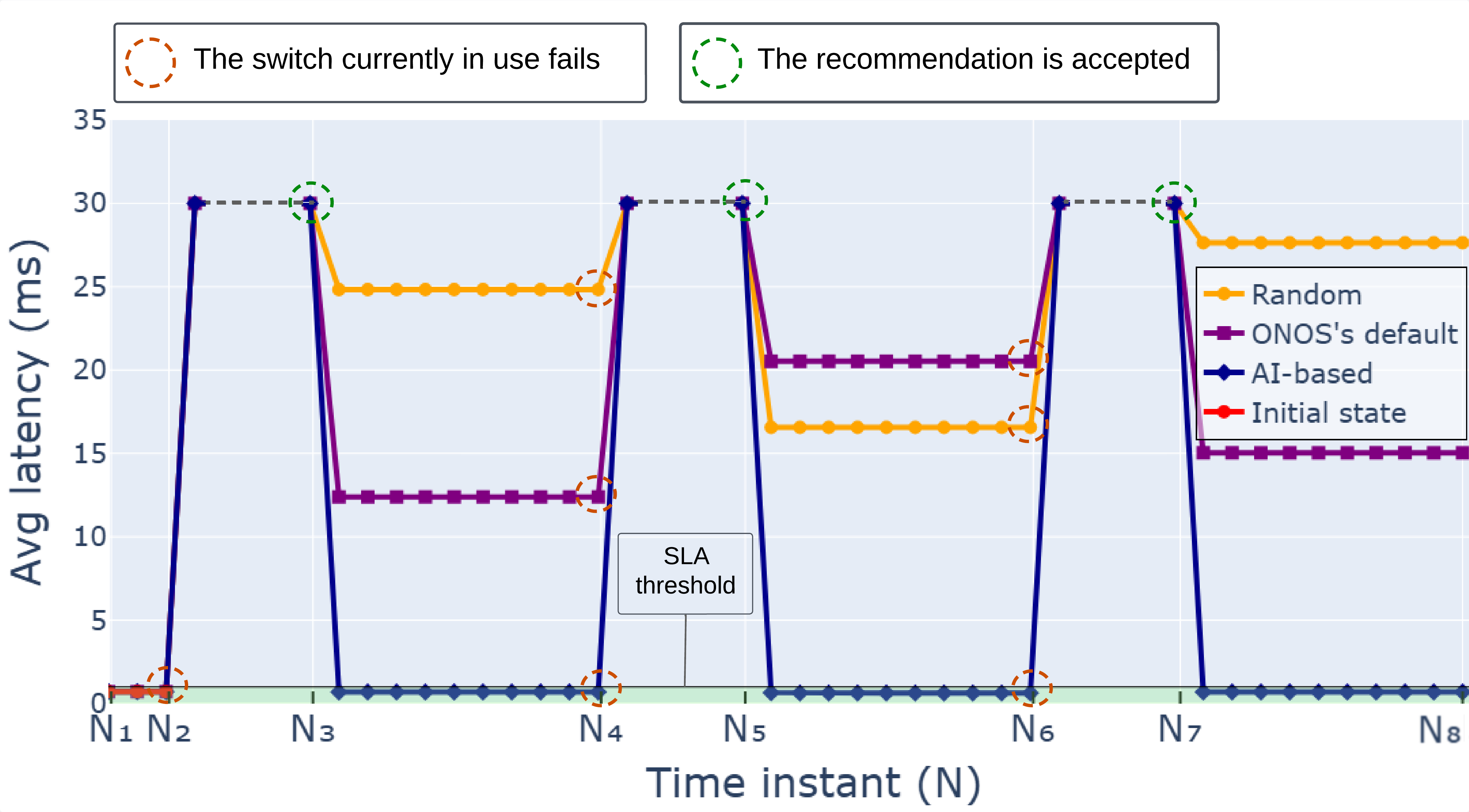}
    \caption{Performance analysis of the URLLC slice in terms of average latency, when applying different routing methods in a 30-node topology.}
    \label{fig:urllc}
\end{figure}

\begin{figure}[ht!]
    \centering
    \includegraphics[width=0.48\textwidth]{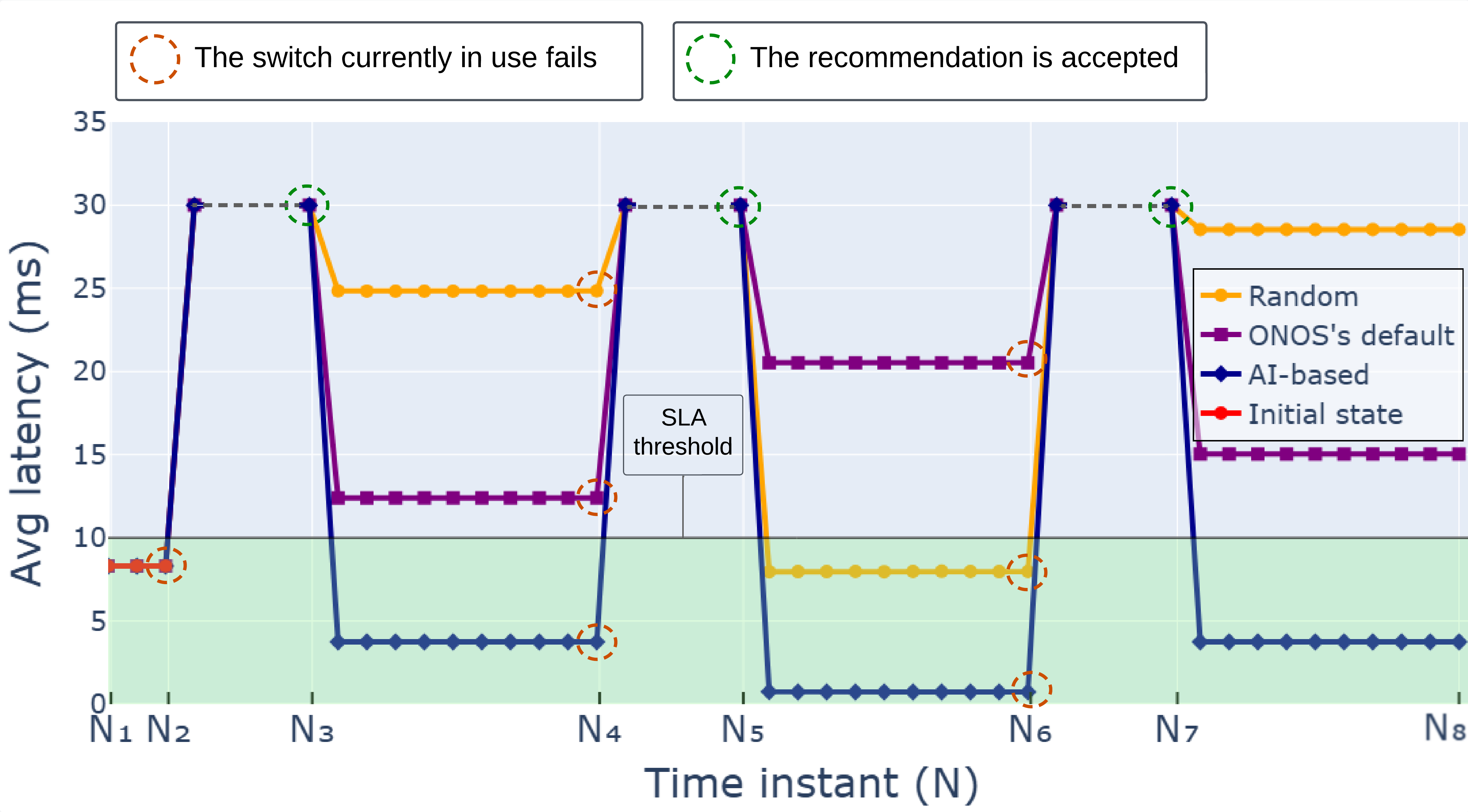}
    \caption{Performance analysis of the eMBB slice in terms of average latency, when applying different routing methods in a 30-node topology.}
    \label{fig:embb}
\end{figure}

In Figs. \ref{fig:urllc} and \ref{fig:embb}, we show how our NDT works, showing the average latency obtained for the URLLC and eMBB slices when applying the recommendation system decisions after three network failures in a 30-node topology. To represent traditional routing methods, we used the default policy of the SDN Controller, which adopts the Open Shortest Path First (OSPF) \cite{ospf}, a link-state routing protocol based on Dijkstra algorithm. This protocol selects the best route based on cost, which, in this case, is the shortest path. 

In the figures, ONOS's default represents the route taken after the switch failure if the NDT was not used. In other words, this route is determined independently of the recommendation system. The experiment begins with the slice allocated on an initial route that meets the latency requirements, represented by the time instant $N_{1}$. At a given moment, defined by $N_{2}$, $N_{4}$, and $N_{6}$, a failure occurs in a switch used by the slice, causing packet loss and increased latency. The average latency of 30 ms represents a very high latency, and the dotted line indicates the period when the slice is without a route. The recommendation system is activated at this point, and the \textit{what-if} analyses are performed. At time instants $N_{3}$, $N_{5}$, and $N_{7}$, a new route is selected to reallocate the affected traffic.

To evaluate the performance of our NDT, we compared the latency predictions generated by the recommendation system with the actual latencies measured on the physical network for each slice for each suggestion solution, in each different node count topologies. For that, we used the Mean Absolute Percentage Error (MAPE), as shown in Table \ref{tab:latency}. The table presents the MAPE values for each solution in different slices and topologies. We achieved a MAPE of approximately 1\% for topologies of 8 and 16 nodes. There was a slight increase to 2\% for the 30-node topology, still ensuring high precision as the network size increases. The NDT-based optimization proved effective in providing valuable insights into changes in the physical network, establishing itself as a powerful tool for testing and supporting intelligent decision-making processes. 

\begin{table}[htp!]
\centering
\caption{Mean Absolute Percentage Error (MAPE) of the predicted end-to-end (E2E) latency for each slice compared to the true latency across 8-, 16-, and 30-node network topologies} 
\scalebox{1.35}{\begin{tabular}{cc|cc}
\cline{3-4}
\multicolumn{2}{c|}{} & \multicolumn{2}{c}{\cellcolor[HTML]{EFEFEF} Routing Algorithm} \\ \hline
\multicolumn{1}{c|}{Topology} & Slice & \multicolumn{1}{c|}{Random} & AI-based \\ \hline 
\hline
\rowcolor[HTML]{EFEFEF} 
\multicolumn{1}{c|}{\cellcolor[HTML]{EFEFEF}} & URLLC & \multicolumn{1}{c|}{\cellcolor[HTML]{EFEFEF}1.73\%} & 1.89\% \\
\rowcolor[HTML]{EFEFEF} 
\multicolumn{1}{c|}{\multirow{-2}{*}{\cellcolor[HTML]{EFEFEF}8 nodes}} & eMBB  & \multicolumn{1}{c|}{\cellcolor[HTML]{EFEFEF}1.84\%} & 1.82\% \\
\multicolumn{1}{c|}{} & URLLC & \multicolumn{1}{c|}{1.92\%} & 1.67\%  \\
\multicolumn{1}{c|}{\multirow{-2}{*}{16 nodes}} & eMBB  & \multicolumn{1}{c|}{1.95\%} & 1.81\%\\
\rowcolor[HTML]{EFEFEF} 
\multicolumn{1}{c|}{\cellcolor[HTML]{EFEFEF}} & URLLC & \multicolumn{1}{c|}{\cellcolor[HTML]{EFEFEF}2.12\%} & 2.28\%\\
\rowcolor[HTML]{EFEFEF} 
\multicolumn{1}{c|}{\multirow{-2}{*}{\cellcolor[HTML]{EFEFEF}30 nodes}} & eMBB  & \multicolumn{1}{c|}{\cellcolor[HTML]{EFEFEF}2.10\%} & 2.15\% \\
\hline
\end{tabular}}
\label{tab:latency}
\end{table}

\section{Conclusion and future work}
\label{sec:conclusions}

This work developed an experimental NDT platform for route optimization in 5G/B5G transport networks, using a recommendation system for \textit{what-if} analyses. The platform effectively generated network performance insights before actual implementation in our physical network, enabling  intelligent decision-making for transport network operations and management. By simulating route changes in different topologies, the virtual twin achieved low MAPE, accurately predicting latencies even in complex scenarios.

Future work will compare route optimization algorithms considering congestion and variable traffic requirements. Response time will also be included in decision-making. To recover SLA's faster, \textit{what-if} analyses will be conducted in advance, enabling pre-stored decisions for critical failures, focusing on slices with high latency demands and reducing storage requirements.

\section*{Acknowledgements}
This work was partially supported by Ericsson Brazil Research and Development Center, by the Project Smart 5G Core And MUltiRAn Integration (SAMURAI)  (MCTIC/CGI.br \newline /FAPESP Grant 2020/05127-2), by Brasil 6G project (RNP/MCTI 
grant 01245.010604/2020-14), and  by the Brazilian National Council for Research and Development (CNPq).
Process for Universal: 405111/2021-5.

\bibliography{reference}

\end{document}